\documentclass{sig-alternate}
\usepackage{colortbl}
\usepackage{url}
\usepackage{verbatim}
\usepackage{footnote}
\makesavenoteenv{tabular}
\usepackage{booktabs}

\begin{document}
%

\title{Analysis of Computer Science Communities \\ Based on DBLP}
\author{
\alignauthor Maria Biryukov\\
\affaddr{University of Luxembourg}\\
\affaddr{FSTC, MINE Group}
\and
\alignauthor Cailing Dong\\
\affaddr{Shandong University}\\
\affaddr{School of Computer Science}
}
\date{08 March 2010}

\maketitle
\begin{abstract}
It is popular nowadays to bring techniques from bibliometrics and scientometrics into the world of digital libraries to analyze the collaboration patterns and explore mechanisms which underlie community development. In this paper we use the DBLP data to investigate the author's scientific career and provide an in-depth exploration of some of the computer science communities. We compare them in terms of productivity, population stability and collaboration trends. Besides we use these features to compare the sets of top-ranked conferences with their lower ranked counterparts.
\end{abstract}

\section{Introduction}
\label{sec:intro}
Computer science is a broad and constantly growing field. It comprises various subareas each of which has its own specialization and characteristic features. At the same time there exist multiple connections between the areas. Thus for example {\it Information Retrieval} combines computer science, linguistics, cognitive psychology, and mathematics. Yet another example, from the area of the {\it World Wide Web}: its rapid growth requires efficient techniques for management of the large volumes of data -- a task that has traditionally been associated with the field of {\it Databases}. The interdisciplinary nature of research is reflected by the conferences' content. Take for instance the {\it Conference on Information and Knowledge Management} (CIKM): besides the topic spelled out in the conference title, it has two other, equally important, streams: {\it information retrieval} and {\it databases}. While different in size and granularity, research areas and conferences can be thought of as scientific communities that bring together specialists sharing similar interests. What is specific about conferences is that in addition to scope, participating scientists and regularity, they are also characterized by level. In each area there is a certain number of commonly agreed upon top ranked venues, and many others -- with the lower rank or unranked. In this work we aim at finding out how the communities represented by different research fields and conferences are evolving and communicating to each other. To answer this question we survey the development of the author career, compare various research areas to each other, and finally, try to identify features that would allow to distinguish between venues of different rank. We believe that such an insight might be of interest for advanced students who are about to choose their specialization; young researchers looking for an appropriate conference to submit their work; authorities who decide on funding of diverse research areas.

This paper is organized as follows: in Section~\ref{sec:related_work} we give an overview of the related work. Section~\ref{sec:data} elaborates on the data collection. In Section~\ref{sec:authors} we discuss the author profiling. Section~\ref{sec:area_compar} focuses on the comparison between various communities and venues. Section~\ref{sec:summary} concludes the paper.
\section{Related Work} 
\label{sec:related_work}

Analysis of large social networks became one of the active research directions in the late $90$-th. Watts and Strogatz~\cite{Watts:Strogatz} contributed to the networks analysis by elaborated discussions on topology, clustering patterns and comparison of random and regular networks. Newman~\cite{Newman:1,Newman:2,Newman:3} has been studying a wide variety of social networks and investigating their essential properties, such as degree distribution, centrality, betweenness, and assortativity, to name a few. The theoretical insight into the principles of social networks yielded a great deal of interest in studying research communities and their properties based on the coauthorship networks.
Nascimento~\cite{Nascimento:2003} has studied network properties of the SIGMOD co-authorship graph. Hiemstra et.al~\cite{Hiemstr:etal} suggested a topological analysis of the Information Retrieval community extracted from the SIGIR records. 
Backstrom, Huttenlocher and Kleinberg~\cite{Kleinberg:etal} have studied mechanisms underlying the membership, growth, and change of the user-defined communities in LiveJournal and DBLP. 
An extensive bibliometric study has been performed by Elmacioglu and Dongwoon Lee~\cite{Elmacioglu:Lee}. Using DBLP to build a co-authorship network they have investigated various properties of the Data Base community and came to the conclusion that DB is a ``small-world" community. Using CiteSeer as a source of bibliograhic records, Huang et. al.~\cite{Huang:etal} applied bibilometric techniques to the analysis of a number of computer science fields in order to study dynamic properties of the underlying netwoks. Based on the top ranked venues recorded in DBLP, Bird et. al~\cite{Bird:etal} identified $14$ computer science communities and studied collaboration patterns and interdisciplinary research at the individual, within-area, and network levels. 

Besides the network property analysis there is an interest in research related to the topic development and distribution in scientific community. 
Z\"{a}ine, Chen and Goebel~\cite{Zain:etal} used collaboration network embedded in DBLP to discover topical connections between the network members and eventually use them in a recommendation system. Another investigation connecting topics and co-authors community has been reported in~\cite{Zhou:etal}. The work used CiteSeer as a testbed and aimed at getting insight in topic evolution and connection between the researchers and topics.

Yet another branch of investigation aims at evaluation of scientific venues. The first attempts relied heavily on the citation networks~\cite{Giles:Council,Sun:Giles:Sidiropoulos}. However as citations are not always available in bibliographic databases other approaches have been proposed. In~\cite{Zhuang:Elmacioglu} criteria for evaluation of program committee members has been developed and successfully applied for ranking conferences recorded in CiteSeer. Yan and Lee~\cite{Yan:Lee} suggested recently a way of ranking venues based on the scientific contribution of individual scholars. The method has been evaluated on ACM and DBLP data sets.

Our work bears on the previous research in that it focuses on statistical investigation of the scientific communities. Its contribution consists in:
\begin{itemize}
\item{extension of a framework for author analysis in order to build a comprehensive profile of the researchers on DBLP;}
\item{setting up and analysis of criteria that allows for both between-area comparison and comparison of conferences that belong to different levels, in an attempt to build up a framework for automatic evaluation of scientific venues.}
\end{itemize} 

\section{Data Collection}
\label{sec:data}

We use computer science bibliographic database DBLP to conduct our investigation. The database is publicly available in XML format at ~\url{http://dblp.uni-trier.de/xml/ }. We downloaded the file in August $2009$ and used conference publications for corpus construction. While DBLP covers $50$ years of publications the data before $1970$ is rather irregular. This is the reason why we consider publications from $1970$ on. 

The complete list accounts for $4449$ distinct conference names. 
Manual examination of the conference pages in DBLP has shown that some venues have changed their names one or more times since they had been established. This observation suggests that we cannot treat conference names as unique because there is no guarantee of capturing the entire history of a venue. Fortunately all instances of the the same conference can be automatically identified with the XML tags in the original file. We use this feature and integrate all events of a venue with multiple names under the name of a component with the longest history. Table~\ref{tab:conf_integr} illustrates the idea. Due to the name unification, the number of conferences is brought down to $2626$. Publications from these conferences constitute the most general data set we use for our experiments. It is denoted {\it CS dataset} and represents the entire DBLP in the context of this paper. 

\begin{savenotes}
\begin{table*}
\centering
\caption{Example of Conference Name Integration}
\label{tab:conf_integr}
{\scriptsize
\begin{tabular}{llcccl} \toprule
\hline
Resulting Name&Individual Names&Time span\\   
\hline\hline
AAAI&Agent Modeling&1\\
&Deep Blue Vs kasparov: the Significance for Artificial Intelligence&1\\
&AAAI Workshop on Intelligent Multimedia Interfaces&1\\
&AAAI/IAAAI, Vol.1&1\\
&AAAI/IAAAI, Vol.2&1\\
&AAAI&17\\
&AAAI/IAAI&5\\                     
\hline
\\\bottomrule
\end{tabular}
}
\vspace{-0.2cm}\end{table*}
\end{savenotes}


As we are interested in a comparative analysis of different scientific communities and venues we have to split the entire set of publications into topical subareas. One of the ways to do so is to specify sets of conferences that correspond to every subarea we want to analyze. Thus we select $14$ subareas~\footnote{While it is widely accepted to treat AI as a separate area we have preferred to decompose it into a few components, such as DMML and NLIR. We admit that these constitute only a subset of the highly interdisciplinary topic of AI.} each of which is represented by a set of relevant top ranked conferences with at least $10$ years time span for the sake of data stability~\footnote{We have had to relax the ``min $10$ years time span" requirement when dealing with conferences in Computational Biology and World Wide Web because these are young areas that have started off at the end of $90$s.}. The idea of relying on the top ranked conferences is inspired by works of~\cite{Bird:etal,Huang:etal,Zhuang:Elmacioglu,Yan:Lee}, and is grounded on the assumption that high quality conferences are clearly defined in terms of topics they cover. While every area has a modest number of commonly agreed upon top ranked venues, the assignment remains subjective. This is the reason why we validate the choice of venues by consulting several hand-made conference ranking  
sources~\cite{cr:cscr,tier:cscr,nanyang:cscr} and considered the estimated venue impact provided by~\cite{cr:impact} . To enable a fair comparison we represent each subarea by the same or nearly the same number of conferences~\footnote{In a few cases renowned conferences with less than $10$ years history have been chosen to maintain consistency of the sets' size.}. 
Table~\ref{tab:subjects} shows the resulting data set which is denoted {\it TOP dataset}. 
\begin{savenotes}
\begin{table*}[t]
\centering
\caption{Research Communities and Corresponding Top Conferences}
\label{tab:subjects}
\resizebox {\textwidth }{!}{
{\scriptsize
\begin{tabular}{|l|c|c|c|}
\hline
Area ID&Abbreviation&Area&Conferences\\
\hline\hline
1&ARCH&Hardware\&Architecture&ASPLOS, DAC, FCCM, HPCA, ICCAD, ISCA, MICRO\\
\hline
2&AT&Algorithm\&Theory&COLT, FOCS, ISSAC, LICS, SCG, SODA, STOC\\
\hline
3&CBIO&Computational Biology&BIBE, CSB, ISMB, RECOMB, WABI\\
\hline
4&CRYPTO&Cryptography&ASIACRYPT, CHES, CRYPTO, EUROCRYPT, FSE, PKC, TCC\\
\hline
5&DB&Data Bases \& Conceptual Modeling&DEXA, EDBT, ER, ICDT, PODS, SIGMOD, VLDB\\
\hline
6&DMML&Data Mining, Data Engineering, Machine Learning&CIKM, ECML, ICDE, ICDM, ICML, KDD, PAKDD\\
\hline
7&DP&Distributed\&Parallel Computing&Euro-par, ICDCS, ICPP, IPDPS, PACT, PODC, PPoPP\\
\hline
8&GV&Graphics\&Computer Vision&CGI, CVPR, ECCV, ICCV, SI3D, SIGGRAPH\\
\hline
9&NET&Networks&ICNP, INFOCOM, LCN, MOBICOM, MOBIHOC, SIGCOMM\\
\hline
10&NLIR&Computational Linguistics, Natural Language Processing, Information Retrieval&ACL, EACL, ECIR, NAACL, SIGIR, SPIRE, TREC\\
\hline
11&PL&Programming Languages&APLAS, CP, ICFP, ICLP, OOPSLA, PLDI, POPL\\
\hline
12&SE&Software Engineering&ASE, CAV, FM/FME, Soft FSE, ICSE, PEPM, TACAS\\
\hline
13&SEC&Security&CCS, CSFW, ESORICS, NDSS, S\&P\\
\hline
14&WWW&World Wide Web&EC-web, ICWE, IEEE/WIC, ISWC, WISE, WWW\\
\hline
\end{tabular} 
}}
\vspace{-0.2cm}
\end{table*} 
\end{savenotes}

As one of our goals is to identify a set of features that would help to distinguish between top and non-top conferences, we need a selection of conferences that do not belong to the set of top ranked venues. Using the same human-made sources  we select $6$ areas with $5$ representative conferences each. They are given in table~\ref{tab:rank_3}, and constitute the {\it NONTOP dataset}. 

\begin{table*}[t]
\centering
\caption{Research Communities and Corresponding Non-Top Conferences}
\label{tab:rank_3}
{\scriptsize
\begin{tabular}{|l|c|c|}
\hline
Abbreviation&Area&Conferences\\
\hline\hline
AT&Algorithms \&Theory&APPROX, ICCS, SOFSEM, TLCA, DLT\\
\hline
CB&Computational Biology \& Medicine&APBC, ICB, ISBRA, CBMS, DILS\\
\hline
DB&Data bases&IDEAS, ABDIS, ADC, WebDB, DOLAP\\
\hline
DM&Data Mining&MLDM, IndCDM, ADMA, KES, IDEAL\\
\hline
SeC&Security \& Cryptography&{\scriptsize{SCN, ISC/ISW, ISPEC, ACISP, WISA}}\\
\hline
WWW&World Wide Web&WEBIST, SAINT, WECWIS, ESWC, ICWE\\
\hline
\end{tabular}
}
\vspace{-0.2cm}
\end{table*}

Note that there are some differences between the two sets in terms of topical partitioning and number of covered subareas. This is explained by the fact that the data about the lower ranked conferences is less consistent and agreeable, and we have preferred to construct smaller though more reliable sets.  

In these three sets above we exclude all publications that have incomplete bibliographic data such as missing authors, title or year. These constitute $0.052\%$ of the records. The remaining publications are used to build co-authorship graphs $G_{CS}$, $G_{Top}$, and $G_{nonTop}$, where $G_{Top}, G_{nonTop} \in G_{CS}$. 
These are undirected graphs where the authors constitute the set of vertices $\{V\}$, and two vertices $v_i, v_k \in \{V\}$ are connected by an edge $e^{'}  \in \{E\}$ 
iff $v_i$ and $v_k$ have coauthored at least one paper. Our experiments are based on these graphs along with other bibliographic data such as number of records, venue, year.

\section{General Researcher Profiling}
\label{sec:authors}

The authors in co-author network are typically investigated from the point of view of their contribution to the research. Thus particular attention is paid to the members of program committees~\cite{Zhuang:Elmacioglu}, ``fathers" of the influential research directions~\cite{Zhou:etal}, authors with high citation index~\cite{find1} or yet those researchers who get often acknowledged~\cite{Giles:Council}. Such an approach yields an interesting but narrow image of the researchers community. In this section we aim at providing a broader view on the authors in entire DBLP and the areas described above by looking at their typical career length, interdisciplinary interests, individual performance pattern and publication distribution with respect to the top and non-top venues. Since our NONTOP dataset covers only a small part of the lower ranked venues listed in DBLP, we do not compare the TOP and NONTOP datasets to each other in this setting. Rather we contrast the data in TOP dataset to the global author statistics in DBLP. 


\subsection{Author career length}
\label{sub:career}  

DBLP contains to hundreds of thousands distinct authors. But how many of them pursue a long scientific career?

\begin{figure}
\centering
\includegraphics[scale=0.68]{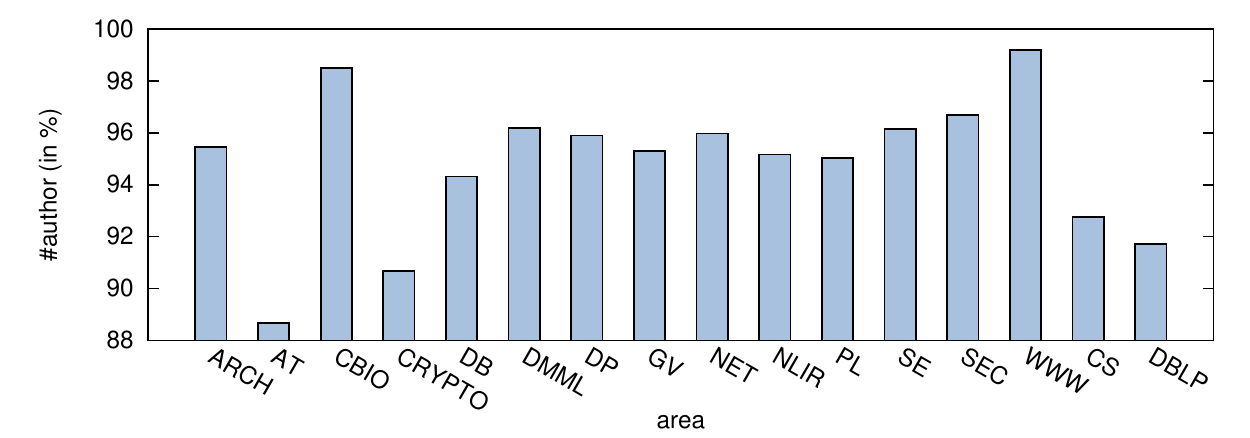}
\caption{Percentage of authors with $\le 5$ years career in TOP set and entire DBLP.}
\label{fig:career_5}
\end{figure}

\begin{figure}
\centering
\includegraphics[width = 0.95\linewidth, height = 0.16\textheight]{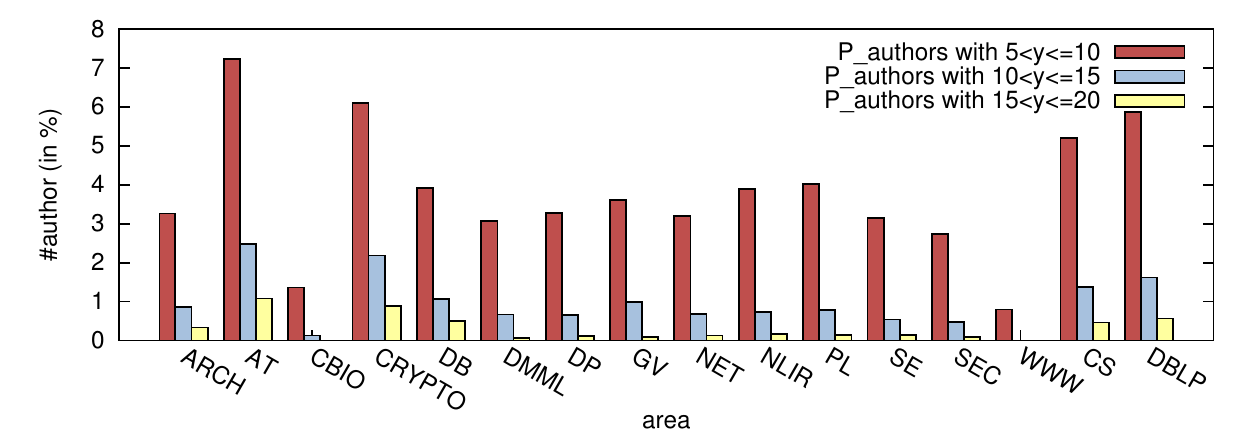}
\caption{Percentage of authors with $6 \le  career \le 20$ years in TOP set and entire DBLP. }
\label{fig:career_rest}
\end{figure}

Figures~\ref{fig:career_5} and ~\ref{fig:career_rest} give a full account on the authors career length distribution among the various research areas in the TOP set, CS dataset, and DBLP as a whole. The first chart represents percentage of authors with $\le 5$ career length, while the second one covers periods from $6$ to $20$ years. 
It turns out that 
top-ranked venues are dominated by authors with $\le 5$ years experience, and only $\approx2\%$ stay publishing at top ranked conferences for more than $10$ years.
This is consistent with the figures obtained on the whole DBLP set: $\approx1.4\%$ of authors have a longer than $10$ years career. We hypothesize that the main component of DBLP authors is represented by PhD students who, after having finished their studies, leave the active scientific career. With respect to the research subareas, AT and CRYPTO have the lowest percentage of researchers with a short career and the highest percentage of people whose career length ranges between $10$ and $15$ years. The explanation lays probably in that fact that these domains require substantial mathematical background and thus time to obtain it which makes them harder to get in for the short time scientists, and more difficult for switching for those who spent so much time on it. 

\subsection{Some characteristics of "experienced"\\ scientists}
\label{sub:publdistr}

We now turn our attention to the authors with $\ge 10$ years experience since they are more probable to influence scientific community than ``short time" researchers. There are $16192$ ($\approx 3\%$) such authors in the whole DBLP set, and $2623$ researchers have $\ge10$ years publication record in the TOP set. 
We characterize 
this latter group in terms of interdisciplinarity of interests and productivity distribution. 
\subsubsection{Interdisciplinarity of Interests}
\label{subsub:interdisciplinarity}

Researchers do not necessarily stay in one and the same field throughout the whole career. But how many areas and at what time of their career do they typically join? What is the  probability for a researcher to join one more area given that he is already publishing in some field.

There are $2623$ authors in the TOP dataset whose career is $\ge 10$ years. Out of them only $\approx 29\%$ work in one area only. The remaining $71\%$ join multiple areas with the average value of $\approx 2.2$. 
We have analyzed the data distribution and found 
that they typically publish in more than one area from the very beginning of the career with a small spike between the 5th and tenth years. It is logical to assume that the interdisciplinarity of the researcher interests serves as an indicator of the area relatedness which can be calculated. For this purpose, let $A_{start}$ be an area in which the author $a_i$ started to publish~\footnote{When calculating the most related areas we assume that an author is publishing in some area iff he has $\ge 2$ publications in it.}.  Next, build a transition matrix $P_{A_i}$ with probabilities $P_{transition} = P_{A_j}|P_{start}$ such that $1 \le j \le 14$, and $j \ne start$. Note that there exist two basic scenarios: $a_i$ publishes in more than one area in one year, and $a_i$ publishes in one area in a given year while overall he is active in multiple areas. We treat these two cases equally when computing $P$. 

\begin{figure}
\centering
\includegraphics[width=0.95\linewidth, height=0.30\textheight]{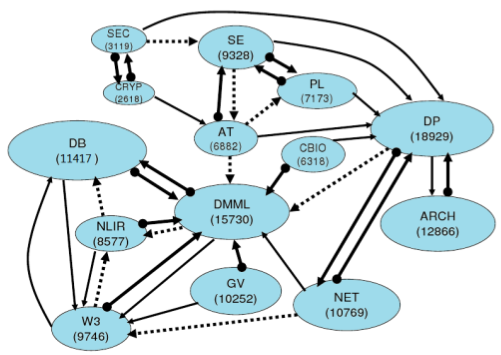}
\caption{Area relatedness based on the researchers' multidisciplinary interests. }
\label{fig:area_relatedness}
\end{figure}

The diagram in Figure~\ref{fig:area_relatedness} shows the most probable transitions between the areas. Each circle represents an area, and its size is defined by the number of people working in it. The thickest  arrows connect the most related areas, the thinner but solid arrows correspond to the second choice and the dotted ones (when present) to the third. The diagram shows clearly that the area relatedness is asymmetric.  For example, Data Mining and Machine Learning (DMML) is primarily related to the Data Bases (DB). At the same time information retrieval (NLIR), computational biology (CBIO), graphics (GV), and WWW have their closest relationship to the DMML, indicating that the authors from these domains publish actively at DMML conferences. It is natural since these more practical areas constitute a field of application for the data mining and machine learning algorithms. 


It is also interesting to note that our rather global results that capture the state of interdisciplinarity in computer science in the last $40$ years, are comparable to the yearly snapshots of the area overlap, found in~\cite{Bird:etal}.   
For example, both claim that there is a considerable authors' overlap between CRYPTO, Security (SEC), and theory (AT); Programming Languages (PL), Software Engineering (SE), and Distributed Computing (DP); Networks (NET) and DP. The similarity of findings that result from static and dynamic computations might point to the long-term relatedness between the areas.   

\subsubsection{Individual Performance Pattern}
\label{subsub:performance}

Let us now focus on the author publication distribution over time and venues. For the temporal distribution analysis we  distinguish between the following three groups of authors:
\begin{itemize}
\item{Authors with $\ge 10$ years experience of publishing in TOPset conferences and focusing on one area only;}
\item{Authors with $\ge 10$ years experience of publishing in TOPset conferences and focusing on multiple areas;}
\item{Authors $\in$ the TOPset with $\ge 10$ years experience of publishing in the CS dataset, irrespective of the number of areas and conference rank.}
\end{itemize}

\begin{figure}[h!]
\centering
\includegraphics[width=0.9\linewidth, height=0.15\textheight]{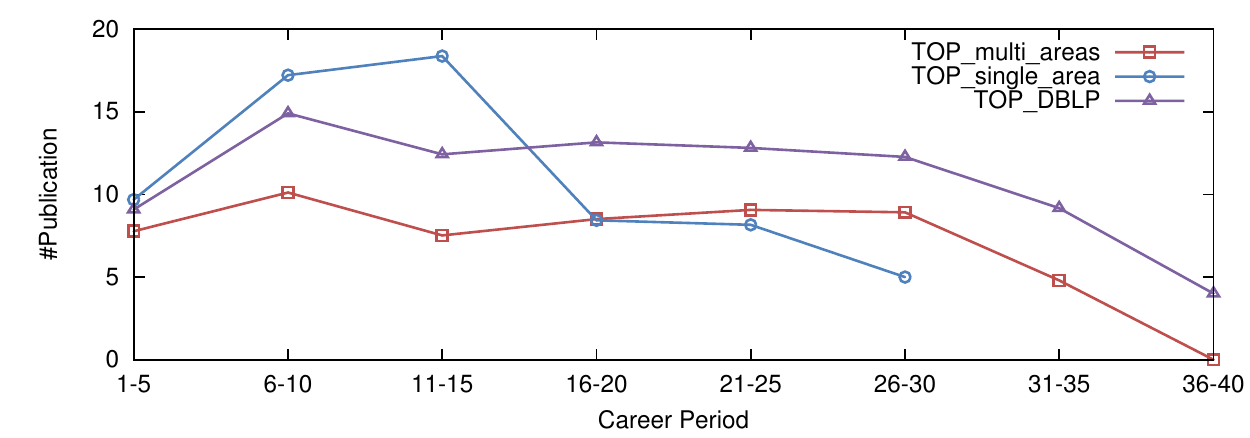}
\caption{Author productivity within the different periods of career.}
\label{fig:performance}
\vspace{-0.1cm}
\end{figure}

The average number of publications produced by each category of authors per $5$-years periods are plotted at Figure~\ref{fig:performance}. 
 The data reveals an interesting pattern: researchers in all three categories are much more active in the $2{nd}$ period of their career, and the single-area authors are even more active in the $3{rd}$ period. After that the productivity drops in the fourth period and remains stable with some minor fluctuations. 
 Based on 
 it we can try to reconstitute the principle milestones in the scientists' life: the first $5$ years correspond roughly to the PhD. studies during which one typically produces a certain (not necessarily high) number of publications. The next $5 - 10$ years ($2$nd period) are of great importance to those who stay in research. In that time authors are evaluated on the international scale and their academic position depends heavily on their productivity. Recall also from the Subsection~\ref{subsub:interdisciplinarity} that the small raise in the number of areas joined by researchers falls into this period, as well.  The later stages correspond to the scientific maturity when scientific output stabilizes on average.

With respect to the publication rate values, they are much higher for the single-area authors during the spike periods.
There is no additional evidence that would help to explain this phenomenon. We might hypothesize that by working in one field only it is easier to get more papers published, since the author knows better the research criteria of his community. 

To analyze the author - publication distribution over venues we calculate for each author $a_i\in$ TOP dataset the percentage of his publications in the top-ranked conferences relative to all his publications recorded in DBLP. Next we combine the results into the $10\%$-intervals and match them against the corresponding percentage of authors. 

\begin{figure}
\centering
\includegraphics[width=1.08\linewidth, height=0.15\textheight]{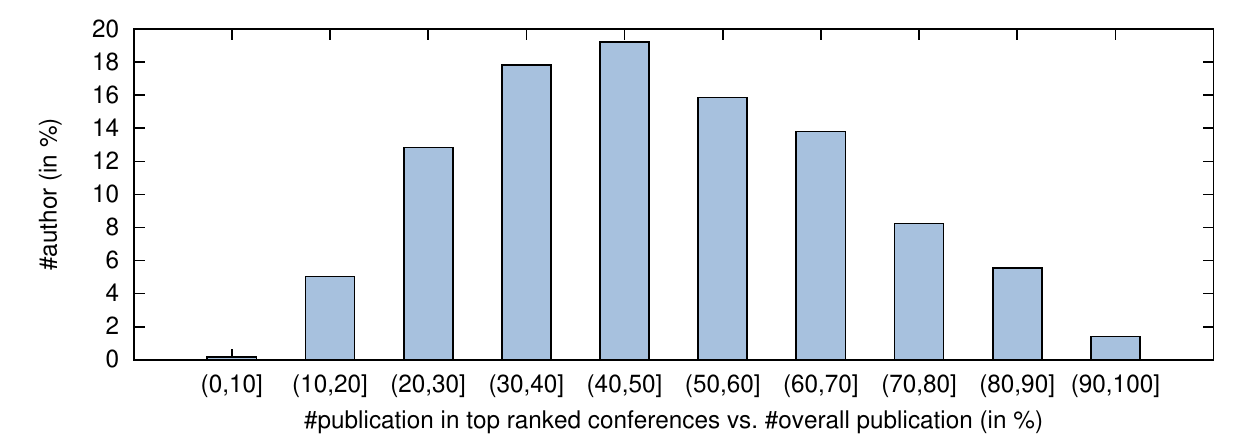}
\caption{Author - venue distribution: percentage of publications at top ranked conferences compared to the overall author production.}
\label{fig:venue_distribution}
\end{figure}

The results are shown at Figure~\ref{fig:venue_distribution}. 
  It turns out that only about $1.5\%$ of authors in the TOP dataset publish exclusively or mostly at the top-ranked venues. Typically the top-ranked conference publications constitute from $30\%$ to $60\%$ of the author's conference production. It suggests that the majority of researchers appears in the mixed set of venues. 

To look closer at the publication distribution over venues in the topical sets we first assign each author $a_i\in$ TOP dataset to the area he contributes at most (frequency based majority voting), and perform the same computation as before~\footnote{CBIO and WWW are not considered as the resulting sets of authors are too small to produce consistent results.}.  

\begin{figure}
\centering
\includegraphics[width=1.08\linewidth, height=0.155\textheight]{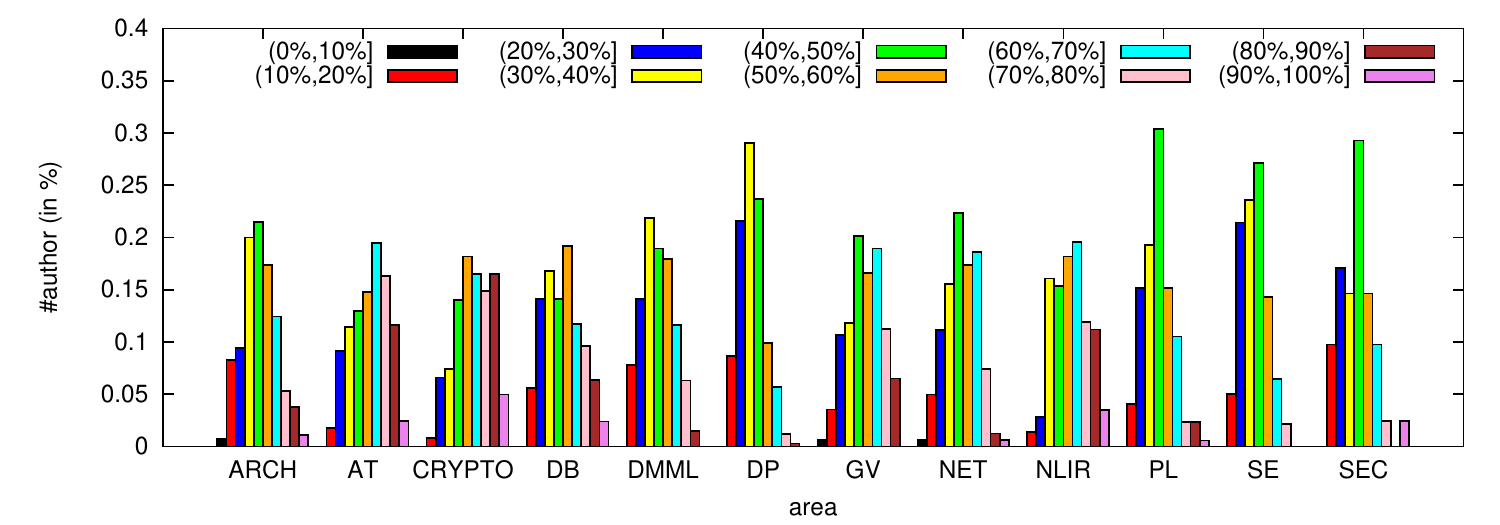}
\caption{Author - venue distribution in various areas.}
\label{fig:area_venue_distr}
\end{figure} 
Figure~\ref{fig:area_venue_distr} presents the results. Notice that majority of areas are dominated by people who publish between $40-50\%$ of their publications in the top ranked conferences, and in DP and DMML the prevailing range is $30-40\%$. These values confirm the general tendency of publishing in the mixed set of venues. On the contrary, authors from DB, CRYPTO, AT and NLIR show more adherence to the top-ranked venues as proportion of researchers who publish $50-70\%$ of papers at top-ranked conferences outranks the other categories.

\section{Scientific Community Analysis}
\label{sec:area_compar}

The previous section dealt with the author characteristic with respect to DBLP and the research areas defined in Section~\ref{sec:data}. In this section we take a closer look at the areas themselves and investigate them in terms of the {\it publication growth rate}, {\it collaboration trends}, and {\it population stability}. Selection of the evaluation criteria is not random. We believe that it may help to highlight the peculiarities of the individual domains and compare them to each other. We apply the same set of features to the subset of the non- top ranked conferences and eventually find out the differences between the top and non-top venues. 

\subsection{Publication Growth Rate}
\label{sub:pub_growth}

Publication growth rate provides an evidence for the area ``well-being" and sheds light on how much interest there is in it at the given moment. It is a dynamic measure that traces yearly changes in the area productivity. We distinguish between the {\it relative} and {\it absolute} growth rates. 

The {\it absolute growth rate} $AbsGr_{A_{i,y}}$ of an area $A_i$ in year $y$ is a ratio of publications in $A_i$ within two consecutive years $y_i$ and $y_{{i-1}}$ such that $AbsGr_{A_{i,y}} = \frac{Publ_{A_{i,y}}}{Publ_{A_{i,y-1}}}$. We have calculated the values for all areas and found that except for the fluctuations corresponding typically to the beginning years, the fields differ considerably from each other. For example, Computer Architecture (ARCH) and Computer Networks (NET) have stabilized at early $90$s, their absolute growths rate values oscillate around $1 \pm 0.1$. On the contrary, Natural Language Processing and Information Retrieval (NLIR) productivity may vary three times as much from year to year, up to nowadays. Such a diversity could probably result from within-venue conventions that define the number of yearly accepted papers. We therefore compare the conferences in our TOP and NONTOP data sets with regard to the absolute publication growth rate. 
It turns out to be systematically higher in the non-top conferences. We can translate this result in terms of {\it publication acceptance rates} (information that is typically not present in the bibliographic databases though it is one of the important parameters for conference evaluation~\cite{Zhuang:Elmacioglu,Yan:Lee}),  and conclude that they are lower for the top venues.   

The {\it relative growth rate} of an area $A_i$ in year $y$, $RGr_{A_i,y}$ is a measure of its activity compared to the overall activity in 
Computer Science (CS)~\footnote{Here, CS is formally represented by either TOP or NONTOP set. However due to the relatively small size of the NONTOP set and the limited number of areas it contains, we rather focus on the TOP set when discussing this metric.}. It is calculated as a ratio between the area absolute growth rate and the computer science absolute growth rate in the given year: $RGr_{A_i,y} = \frac{AbsGr_{A_{i,y}}}{AbsGr_{CS_{i,y}}}$
Thus $RGr_{A_i, y} > 1$, indicates a raise of interest to the area $A_i$ in some year $y$. 

\begin{figure}
\centering
\includegraphics[scale=0.68]{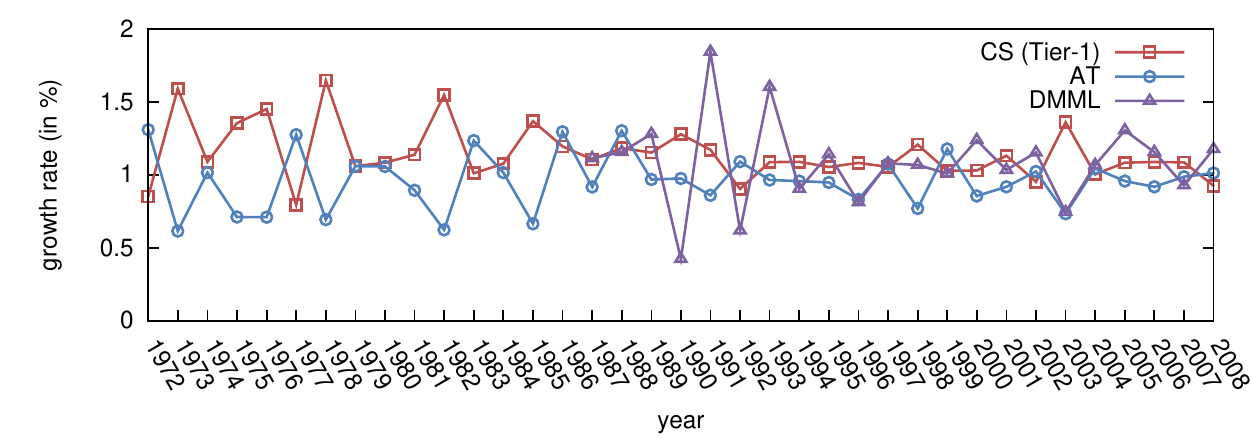}
\caption{Relative growth rates of AT and DMML vs absolute growth rate of CS}
\label{fig:relative_growth}
\end{figure}

Figure~\ref{fig:relative_growth} illustrates the idea.
As of CS, we observe considerable fluctuations in its growth rate with the overall tendency to raise in the $70$s - 1{st} half  of $80$s. One possible explanation is that many areas had started off  in that period. At the same time the diapason in conference productivity is large in the beginning, and this is the reason why the curve goes up and down rather than increasing steadily. An additional explanation of the unstable behavior of the curve is the incompleteness of the DBLP data for the corresponding period. On the contrary, influx of the new disciplines becomes much smaller from the 2{nd} half of the $80$s on, and we notice only two modest spikes - at the end of $90$s and in the first years of $2000$ which reflect most probably the contribution of the new-born Computational Biology, and World Wide Web. 
  
We chose DMML and AT to visualize the concept of the relative growth rate.  
On the background of the global development of CS, the bursts of activity in DMML can be seen in the beginning of $90$s, and several times in the $2000$s, though on the smaller rate. It corresponds well to the evolution of the area which has become very popular in the late $80$s - beginning of $90$s and attracts a great deal of attention nowadays. On the contrary, relative growth rate in AT remains most of the time bellow one. We suppose that the same considerations that we have mentioned in Subsection~\ref{sub:career} prevent the area becoming ``trendy".

\subsection{Collaboration trends}
\label{sub:collaboration}

Analysis of collaborations shows how much community is connected. One might expect that a highly interdisciplinary area such as {\it Data Mining} will exhibit lower connectivity than for example {\it Information Retrieval} which is focused on a much smaller number of topics and thus facilitates the collaboration. In addition to the between-area comparison we investigate the difference in collaboration pattern in communities described by the TOP and NONTOP data sets. 
The collaboration pattern is analyzed in terms of an average number of coauthors per paper and per author, and clustering coefficient which quantifies how close the direct neighbors of a vertex are to form a complete graph~\cite{Watts:Strogatz}. We use co-authorship graphs  $G_{CS}$, $G_{Top}$, and $G_{nonTop}$ defined in Section~\ref{sec:data} along with publication statistics to perform these computations.

Previous analysis of the co-author network in ACM data set has shown that the number of collaborators per author increases steadily over the years~\cite{Zhuang:Elmacioglu}. It has been confirmed by~\cite{Huang:etal} who used CiteSeer as the experimental testbed. Our results obtained from the DBLP show that the increasing average number of co-authors per authors as well as the average number of authors per paper characterize all the subareas we deal with.   
Tables~\ref{tab:collaborations_top} and~\ref{tab:collaborations_nontop} summarizes our findings.    
\begin{savenotes}
\begin{table*}[ht]
\centering
\caption{Collaboration trends in TOP set}
\label{tab:collaborations_top}
\resizebox {\textwidth }{!}{
{\scriptsize
\begin{tabular}{|c|c|c|c|c|c|c|c|c|}\hline
Area&Vertexes&\# of Authors per paper\footnote{In Tables~\ref{tab:collaborations_top},~\ref{tab:collaborations_nontop}, the average number of authors per paper is given by the tuple $\langle 1^{st}$ year of an area, 2009$\rangle$.}&\# of Coauthors per Author&\# of Coauthors per Author&\# of Coauthors per Author&\# of Coauthors per Author&\% of Singletons&CC\\
&&&in the same area&in the same area&in TOP set&in DBLP (CS set)&&\\
&&&$1^{st}$ year&average over the entire period&&&&\\
\hline
ARCH&12866&1.68-3.05&1.05&5.6&7.9&16.62&3.2&0.71\\
\hline
AT&6882&1.22-2.22&0.48&4.8&9.1&18.3& 8.3&0.5\\
\hline
CBIO&6318&2.80-3.36&2.43&4.9&7.8&15.6&2.3&0.79\\
\hline
CRYPTO&2618&1.50-2.27&1.07&4.61&8.09&15.4&7.3&0.57\\
\hline
DB&28273&1.36-2.70&0.6&4.95&8.87&19.1&5.7&0.68\\
\hline
DMML&15730&2.13-2.80&2.9&4.13&8.34&19.35&3.2&0.67\\\hline
DP&18929&2.05-2.80&1.5&4.36&7.77&19.04&3.1&0.66\\\hline
GV&10252&2.64-3.02&3.85&4.26&5.8&16.9&2.5&0.67\\\hline
NET&10769&1.94-2.84&1.25&3.98&6.93&17.61&2.4&0.66\\\hline
NLIR&8577&1.56-2.63&1.52&4.71&7.45&16.74&7.0&0.66\\\hline
PL&7173&1.77-2.35&1.55&3.74&8.01&18.3&8.7&0.61\\\hline
SE&9328&1.90-2.54&1.5&3.83&7.29&18.64&7.3&0.64\\\hline
SEC&3119&1.52-2.62&0.72&3.7&9.36&21.21&6.1&0.68\\\hline
WWW&9746&2.79-3.12&2.71&3.75&8.01&21.58&2.4&0.74\\
\hline
\end{tabular}
}}
\vspace{-0.2cm}
\end{table*}
\end{savenotes}

\begin{table*}[t]
\centering
\caption{Collaboration trends in NONTOP set}
\label{tab:collaborations_nontop}
\resizebox {\textwidth }{!}{
{\scriptsize
\begin{tabular}{|c|c|c|c|c|c|c|c|c|}\hline
Area&Vertexes&\# of Authors per paper&\# of Coauthors per Author&\# of Coauthors per Author&\# of Coauthors per Author&\# of Coauthors per Author&\% of Singletons&CC\\
&&&in the same area&in the same area&in NONTOP set&in DBLP (CS set)&&\\
&&&$1^{st}$ year&average over the entire period&&&&\\
\hline
DB&2893&2.03-2.39&2.0&2.84&3.7&7.9&5.2&0.62\\
\hline
AT&2761&1.66-1.97&0.95&2.17&3.5&8.09&14.9&0.55\\
\hline
CBIO&4866&2.90-3.26&3.07&4.3&4.7&8.87&2.0&0.83\\
\hline
DM&9434&2.53-2.87&2.43&3.22&3.57&8.09&3.2&0.71\\\hline
SEC&1727&2.07-3.01&1.58&3.34&3.78&8.34&3.4&0.69\\\hline
WWW&6205&2.38-3.04&2.17&3.73&4.2&9.13&3.8&0.75\\
\hline
\end{tabular}
}}
\vspace{-0.2cm}
\end{table*}

In the TOP set data,  
CBIO and WWW have the highest average number of authors per paper along with the highest clustering coefficient which implies intensive collaborations throughout the entire community. On the contrary, AT, CRYPTO and PL (Programming Languages) have the smallest number of authors per paper, highest percentage of singleton authors and the lowest clustering coefficient among all $14$ disciplines. It follows that in these three areas authors have a strong preference for working in small groups when collaborating. Moreover these groups turn to be weakly connected which results in a network composed of rather isolated cliques. 
  It is worth mentioning that~\cite{Bird:etal} found that among other CS areas, CRYPTO has the highest collaborative {\it assortativity}. Assortativity~\cite{Newman:28} quantifies how much a vertex in the network is connected to alike vertices. {\it Collaborative} assortativity reflects the tendency of authors to collaborate with those authors who have similar number of coauthors. This selectivity in collaboration pattern scales well with our assumption about sparseness of the cryptographic community. A bit surprisingly but the figures in the table do not confirm our assumption about the connectivity of DMML and NLIR. The higher percentage of coauthors per author coming from the same area ($63\%$) in NLIR proves its lower interdisciplinarity compared to DMML where $\approx 51\%$ coauthors per author belong to other disciplines. However it does not seem to have an impact on the connectivity pattern, and the clustering coefficient of NLIR is a little smaller than that of DMML. Alternatively it can be explained by the fraction of working alone authors (singletons) which is almost twice as much in NLIR as in DMML and naturally lows down the connectivity rate of the former. The weak relation between the interdisciplinarity of a field and its connectivity is best seen with \{GV (Graphics), SEC (Security)\} pair. The clustering coefficient of both is slightly above average ($0.67$ and $0.68$ vs $0.65$). At the same time GV is the most homogeneous area out of all $14$ ($73\%$ of coauthors per authors belong to GV), while SEC is the most heterogeneous one: only $40\%$ of coauthors per authors come from the same discipline. 
  
The data in Table~\ref{tab:collaborations_top} reveals that on average only $43\%$ of coauthors per author belong to the set of authors publishing at top ranked conferences. It is in line with the author/venue distribution discussed in Subsection~\ref{subsub:performance}, and confirms that the same researchers publish at top and non-top ranked venues. In general, the NONTOP set (Table~\ref{tab:collaborations_nontop}) is featured by the slightly higher number of authors per paper and higher clustering coefficient (DB is an exception), although the values are close in both sets. Note also that  if we were to sort the areas by the clustering coefficient, the order would be the same as in the TOP set (DB and DMML switched around). However we have no sufficient evidence to conclude whether or not the non-top ranked conferences exhibit distinctive behavior in this setting compared to the top-ranked venues.

\subsection{Population Stability}
\label{sub:stability}

In Section~\ref{sec:authors} we discussed area interdisciplinarity as suggested by author transitions between the fields. In this section we concentrate on the mechanism that influence researcher dynamics. For this we analyze changes in conference populations in terms of new members that join a venue ({~\it newcomers}), and those who leave it, {\it  leavers}. In the context of this section, the large {\it communities} corresponding to the research areas are decomposed into the conferences each of which is understood as an individual community.   

In~\cite{Kleinberg:etal} it has been pointed out that the membership in a community may be influenced by fact of having ``friends" in that community. Thus some researchers are more likely to submit their paper to a conference if they have previously coauthored with someone who had already published over there. The theory has been tested on LiveJournal and DBLP (set of $84$ conferences with at least $15$ years history) communities. We take on this approach and investigate whether this property holds equally in different areas and venues. We therefore define:
\begin{itemize}
\item{Newcomer $New_{c_{k,y}}$}: an author who had no publications at conference $c_k$ before year $y$. We define a fraction of newcomers in a conference $c_k$ in the year $y$ as $NewComers_{c_{k,y}} = \frac{\sum{New_{c_{k,y}}}}{TotalAauthors_{c_{k,y}}}$;  
\item{Pure newcomer $Pnew_{c_{k,y}}$}: an author who had neither publications nor has he coauthored with an author already member of $c_k$ before year $y$. The pure newcomers are calculated as $PnewComers = \frac{\sum{Pnew_{c_{k,y}}}}{NewComers_{c_{k,y}}}$;
\item{Leaver $Leaver_{c_{k,y}}$: an author who has no more publications in $c_k$ after year $y$. The fraction of leavers in $c_{k,y}$ is formalized as $\frac{\sum{Leaver_{c_{k,y}}}}{TotalAauthors_{c_{k,y}}}$}.
\end{itemize}

Results of the computations are given in Tables~\ref{tab:stability_top},~\ref{tab:stability_nontop}. Due tot he space considerations we show only the most interesting results.
\begin{table*}[ht]
\centering
\caption{Population stability in TOP set}
\label{tab:stability_top}
{\tiny
\begin{tabular}{|c|c|c|c|c|l|}\hline
Area&Conference&$1^{st}$ year&Average NewComers&Average PnewComers&Average Leavers\\\hline
ARCH&FCCM&1995&0.72&0.53&0.70\\
&HPCA&1995&0.65&0.44&0.63\\
&ICCAD&1990&0.56&0.31&0.54\\
&ISCA&1973&0.64&0.45&0.59\\
&MICRO&1987&0.63&0.44&0.59\\
&ASPLOS&1982&0.78&0.56&0.74\\
&DAC&1985&0.61&0.38&0.57\\\hline
AT&FOCS&1970&0.48&0.44&0.41\\
&ISSAC&1988&0.49&0.57&0.48\\
&LICS&1986&0.53&0.54&0.51\\
&SODA&1990&0.51&0.39&0.42\\
&STOC&1970&0.44&0.43&0.38\\
&COLT&1988&0.44&0.48&0.40\\
&SCG&1986&0.45&0.32&0.41\\\hline
CRYPTO&EUROCRYPT&1982&0.48&0.45&0.46\\
&FSE&1993&0.50&0.47&0.46\\
&ASIACRYPT&1990&0.60&0.56&0.58\\
&CHES&1990&0.64&0.63&0.64\\
&CRYPTO&1981&0.47&0.45&0.46\\
&PKC&1998&0.63&0.53&0.61\\
&TCC&2004&0.52&0.29&0.49\\\hline
DMML&ECML&1987&0.74&0.72&0.64\\
&ICDE&1984&0.63&0.44&0.55\\
&ICML&1988&0.60&0.51&0.52\\
&KDD&1994&0.67&0.53&0.59\\
&PAKDD&1998&0.74&0.67&0.68\\
&CIKM&1992&0.76&0.65&0.68\\
&ICDM&2001&0.75&0.66&0.69\\\hline
NLIR&EACL&1983&0.82&0.8&0.76\\
&ECIR&1997&0.76&0.7&0.65\\
&ACL&1979&0.66&0.64&0.52\\
&SIGIR&1971&0.64&0.63&0.55\\
&SPIRE&1998&0.67&0.66&0.65\\
&TREC&1992&0.49&0.40&0.43\\
&NAACL&2001&0.74&0.59&0.61\\\hline
SEC&ESORICS&1990&0.77&0.69&0.72\\
&NDSS&1997&0.78&0.64&0.75\\
&CCS&1993&0.73&0.61&0.58\\
&CSFW&1988&0.55&0.59&0.50\\
&S\&P&1980&0.75&0.65&0.70\\\hline
WWW&ISWC&1997&0.70&0.57&0.68\\
&EC-web&2000&0.82&0.80&0.85\\
&ICWE&2003&0.71&0.73&0.76\\
&IEEE\/WIC&2001&0.82&0.79&0.79\\
&WWW&2001&0.73&0.58&0.70\\
&WISE&2000&0.83&0.75&0.83\\\hline
\end{tabular}
}
\end{table*} 

\begin{table*}[t]
\centering
\caption{Population stability in NONTOP set}
\label{tab:stability_nontop}
{\tiny
\begin{tabular}{|c|c|c|c|c|l|}\hline
Area&Conference&$1^{st}$ year&Average NewComers&Average PnewComers&Average Leavers\\\hline
AT&APPROX&1998&0.75&0.72&0.64\\
&ICCS&1992&0.53&0.59&0.48\\
&SOFSEM&1995&0.82&0.83&0.79\\
&TLCA&1993&0.66&0.74&0.65\\
&DLT&1993&0.56&0.66&0.54\\\hline
DM&MLDM&1999&0.85&0.86&0.75\\
&IndCDM&2001&0.86&0.84&0.75\\
&ADMA&2005&0.85&0.75&0.87\\
&KES&1997&0.79&0.75&0.75\\
&IDEAL&2000&0.81&0.75&0.81\\\hline
SEC&SCN&2002&0.75&0.71&0.74\\
&ISC\/ISW&1997&0.83&0.75&0.83\\
&ISPEC&2005&0.65&0.58&0.84\\
&ACISP&1996&0.86&0.76&0.62\\
&WISA&2003&0.84&0.79&0.87\\\hline
WWW&WEBIST&2005&0.89&0.90&0.88\\
&SAINT&2001&0.81&0.72&0.78\\
&WECWIS&1999&0.84&0.81&0.81\\
&ESWC&2004&0.75&0.62&0.69\\
&ICWE&2003&0.71&0.73&0.76\\\hline

\end{tabular}
}
\end{table*} 

Let us discuss some of the TOP set conferences. All venues in AT and CRYPTO prove stable and moreover are the most stable venues in the whole TOP set. They are characterized by low percentage of Newcomers, Pure newcomers, and Leavers, compared to the average values across the whole TOP set.  Note that fraction of Pure newcomers is an important parameter as it sheds light on how ``friendship" phenomenon affects the inflow of the new authors: the higher the fraction is, the smaller is the friendship influence. We have found that AT and CRYPTO are friendship driven as about $50\%$ of new authors joining venues have co-authored with authors who had already published over there. 

Contrarily to the two fields above, WWW conferences are the most dynamic ones, featured by the high values for the Newcomers, Pure newcomers, and Leavers' fractions.       
Friendship does not seem to alter the influx of new authors as the Pure newcomers typically count for $\approx 60-80\%$ of all the Newcomers. Note that the member conferences are young - except of ISWC that has started off in $1997$ all other venues have appeared in $2000$s. It is natural to postulate that the population stability of a venue is directly related to its age. In the given set of conferences, our assumption is immediately confirmed by the ISWC which has the lowest values for all three aspects. Note however that the above relation holds in many but not all the cases. Thus for example in Security, CSFW ($1988$) is less dynamic than S\&P ($1980$), and ICCAD ($1990$), the most stable community in Architecture, is much younger than ISCA ($1973$) which scores second in terms of stability. The interpretation of these observations is that while population stability does depend to the certain extent on the conference age, it is also influenced by other, conference specific factors.     

The key observation concerning the NONTOP set of venues, is that all of them irrespective of time span (which ranges from $17$ to $3$ years) and domain, are very dynamic. (The only exceptions are ICCS and DLT (AT) whose behavior is closer to AT venues from the TOP set). Typically the Newcomers constitute about $75 - 85\%$ of all authors, and the average value of the Pure newcomers is about $75\%$ which suggests that the friendship influence on the decision to join a venue is rather negligible. The turnover of authors is also remarkable since the fraction of Leavers is often comparable to that of Newcomers and constitutes up to $88\%$ of all the authors. As such, population stability might be considered as a candidate feature that helps to distinguish between the top and non-top venues. 

\bigskip

\section{Conclusions and Future Work}
\label{sec:summary}

In this paper we have analyzed computer science communities in different settings. We a performed statistical analysis of authors, and found that the DBLP community is dominated by the short-time researchers whose career does not exceed $5$ years. We have also discovered that experienced scientists from the top-ranked venues tend to join multiple research communities and produce the highest number of publications between the 5{th} and 10{th} years of their career. Typically they publish in a mixture of top and non-top ranked venues.  

We have also compared communities from $14$ research areas of computer science and performed the between-area comparison in terms of publication growth rate, collaboration trends and population stability. In addition, we applied the same criteria to the comparison between top and non-top ranked conferences and discovered that the publication growth rate and population stability could be among the features that help to separate the two sets. 

In this approach we have manually divided the broad area of computer science into $14$ topics. In the future we plan to substitute this rather ad hoc approach by applying a machine learning technique such as {\it Latent Dirichlet Allocation}~\cite{Blei:2003} for both - topic classification and learning the best number of topics into which the given data can be divided. By doing this we will avoid the subjectivity of manual classification.    
We also plan to elaborate on the set of features that could be used for efficient comparison and eventually automatic ranking of venues. Besides we plan to extend the notion of ``venue" to incorporate journals into analysis. 

\section{Acknowledgments}

We would like to thank Prof. Christoph Schommer for his critical reading, valuable comments and helpful suggestions.  

\bibliographystyle{abbrv}
\bibliography{ecdl_v03}  
\end{document}